\documentstyle[aps,amssymb]{revtex}
\begin{document}
\title{Is the Universe a Quantum System?}
\author{Helmut Fink and Hajo Leschke\\ \vspace{0.5cm}
{\small 
Institut f\"ur Theoretische Physik \\
Universit\"at Erlangen-N\"urnberg\\
Staudtstr.\ 7\\91058 Erlangen, Germany}}
\maketitle

\begin{abstract}
In order to relate the probabilistic predictions of quantum theory 
uniquely to measurement results,
one has to conceive of an ensemble of identically prepared copies of the
quantum system under study. Since the universe is the total domain of physical
experience, it cannot be copied, not even in a thought experiment.
Therefore, a quantum state of the whole universe can never be made 
accessible to empirical test. Hence the existence of such a state
is only a metaphysical idea. 
Despite prominent claims to the contrary, recent developments in the
quantum-interpretation debate do not invalidate this conclusion.
\end{abstract}

\section{Introduction}

A hundred years after Planck's quantum hypothesis, 
quantum theory seems to be universally valid. While it has its roots 
in the atomic and subatomic domain, the stability of
matter, diamagnetism
and superconductivity are examples of quantum effects in the macroscopic
domain.
A fundamental limitation on the applicability of quantum theory has not 
been accepted so far. The formalism of general quantum 
theory allows one to incorporate into a single quantum description 
additional degrees of freedom as a subsystem, for example a heat bath
or a detector in a laboratory. 
The formalism contains no fundamental obstacle to the description of
an arbitrary
compound of physical microsystems. Thus it is tempting to think even of
the universe as a whole in terms of quantum theory \cite{PI86}. 
Of course, nobody
can explicitly specify a quantum state of the universe \cite{HH83}. 
Nevertheless, assuming its existence
in principle is enough to construct theoretical models
of a quantum universe and to study their predictions, at least at a formal 
level. Demanding research programs, such as quantum gravity and quantum
cosmology \cite{HI87,GH93,Lin98}, partially rest on this idea.

In spite of these efforts, in
this article we will demonstrate that the very concept of a 
``quantum state of the universe'' is doomed to failure.
While this conclusion is not entirely new \cite{Lud85}, 
we feel that it has not received the attention it deserves. Indeed,
publications employing a quantum state of the universe, in whatever 
specification, continue to appear. Therefore, our goal here is to reconsider
this concept and clearly present arguments decisive for its rejection.
In this way we also declare our position in the recently revived debate 
on the meaning of quantum theory.

The logical structure of our reasoning is as follows. Taking 
{\bf (U)}, {\bf (F)} and {\bf (MI)} for granted, 
it follows that {\bf (QU)} is excluded. Here
{\bf (U)} stands for a definition of the concept of ``universe'',
{\bf (F)} for the principle that the interpretation of any 
physical theory has to rely on facts,
{\bf (MI)} for a minimal interpretation of quantum theory, and {\bf (QU)} for 
the claim that there is a quantum state of the universe.
Those who tend to escape our conclusion have to decide which of our
assumptions they regard as closest to dispensable.

The article is organized as follows. In section \ref{prerequisites} 
we specify explicitly {\bf (U)}, {\bf (F)} and {\bf (MI)}. 
Section \ref{reasoning} contains our reasoning
against {\bf (QU)}. In the last section, we discuss and refute possible 
objections in an interplay of questions and answers. 

\section{Prerequisites}
\label{prerequisites}

We start with the specification of {\bf (U)}, which defines the most extended
object of physical description.

\vspace{0.5cm}
\noindent
\fbox{\parbox{17cm}{
{\bf (U) Definition of the Universe}. The (physical) {\em universe}
is the union of all objects and phenomena which are empirically accessible in 
principle.}}
\vspace{0.5cm}

Two explanations are in order.
\begin{enumerate}
\item An object or phenomenon is {\em empirically accessible} only if it is
intersubjectively perceptible and communicable. Here it is irrelevant whether
a living being is actually observing the object or phenomenon. 
All that matters is that an observation is possible at any time. 
Empirical accessibility may well require sophisticated 
experimental manipulations or  
technical means 
for observation.
\item ``{\em In principle}'' means 
``supposing perfect measurement apparatuses''. 
What is to be considered as empirically accessible in principle does not  
depend on the current state of measurement-device technique.
In this sense, cosmic background radiation was ``empirically accessible in 
principle'' already in antiquity. In contrast, if quantum theory is true,
simultaneous values of mutually incompatible observables are not 
empirically accessible in principle.
\end{enumerate}

To give some motivation for the principle {\bf (F)}, we first 
recall the purpose 
of interpretations of physical theories in general. The goal of an 
interpretation is a unique relation between the mathematical formalism of the 
theory and the objects and phenomena which are to be described.
Roughly speaking, an interpretation of a physical theory is a set of 
mapping principles relating certain elements of the mathematical formalism 
to certain elements of physical reality. 
If one knows the objects and phenomena to be described, the interpretation 
shows how to apply the formalism. Vice versa, if one knows the formalism, 
the interpretation shows which objects and 
phenomena the theory is able to describe.

\vspace{0.5cm}
\noindent
\fbox{\parbox{17cm}{{\bf (F) Principle of Relation to Facts.} Every 
interpretation of a physical theory has to relate certain elements of the 
mathematical formalism of the theory to {\em conceivable facts}.}}

\vspace{0.5cm}
Some 
explanations are in order.
\begin{enumerate}
\item Here a {\em fact} is only what is 
empirically accessible in principle in a single measurement on an 
individual system. 
\item {\em Conceivable} facts are facts that {\em may} but need not exist 
in reality, supposing the 
theory is exactly valid. They are the kind of facts considered in 
thought experiments. 
We note that in some situations it is
empirically clear that they do not exist in reality, as is the case in 
counterfactual reasoning \cite{Sta97}.
\item The purpose of the concept of {\em conceivable} facts is not to test the 
empirical adequacy of the theory (for that the {\em real} facts are decisive), 
but only to specify the testible statements of the theory. 
Strictly speaking, the theory can only 
be tested empirically when its interpretation has been specified by means 
of conceivable facts. In this sense, interpretation is a precondition 
of testability.
\item Principle {\bf (F)} excludes an understanding of ``interpretation''
in the broad sense of attributing ``meaning'' to a theoretical concept
by means of free human imagination regardless of any empirical relevance.
Speculative imagination in physics, useful as it is for the invention of
new hypotheses, has to pay tribute to the methodological basis of 
theoretical concepts, which is epitomized in principle {\bf (F)}.
\end{enumerate}

Examples of conceivable facts are the values of all observables in classical 
mechanics. A probability density on phase space is not a conceivable fact, 
since
it incorporates some ignorance about the classical system under consideration.
However, the point in phase space that describes the ``real'' state
of the system represents a conceivable fact, even if it is not precisely known
(which is the typical situation in classical statistical mechanics).

In quantum theory, all internal parameters that characterize 
a quantum system (such as 
mass, spin, charge etc.) stand for conceivable facts. 
In contrast, a value of a quantum observable (such as position, 
energy, orbital angular momentum etc.)
is never assigned as a fact to a system in all of its states 
that show quantum uncertainty for this observable, namely in its
non-eigenstates.
Also, it is not a conceivable fact that a certain
Schr\"odinger wave function (in other words, pure state) is given, 
since this function 
cannot be tested by a single measurement on an individual system.
Hidden variables would be connected with conceivable facts (hence the efforts
to introduce them), but they are not part of the quantum formalism.
In any case, results of measurements performed on an individual 
quantum system are always conceivable facts. 

Our third premise is the set of basic rules of how to apply 
quantum theory. These rules are almost uncontroversial 
among the proponents of different quantum interpretations.

\vspace{0.5cm}
\noindent
\fbox{\parbox{17cm}{{\bf (MI) Minimal Interpretation of Quantum 
Theory.} Every state of a given quantum system yields probabilistic
predictions for all observables that can be measured on this system.
More precisely, let
the quantum state $\cal W$ be represented by a positive trace-one 
operator $W$ acting on some complex separable
Hilbert space $\cal H$, and the observable $\cal A$ by a 
positive-operator-valued measure $E_{\cal A}$ on a suitable set $\Omega $ 
with measurable subsets $X \subseteq \Omega $. 
Then the trace
${\rm Tr}\left[ WE_{\cal A}(X)\right] $ 
is the {\em probability} of finding a result in 
$X$ when $\cal A$ is measured on the system in the state 
$\cal W$. }}

\vspace{0.5cm}
Some comments apply.
\begin{enumerate}
\item Remarkably, the possible outcomes of measurements in the sense of
experimental physics are related to the notions ``measure'' and
``measurable subset'' in the sense of mathematical measure theory 
(see, for example
\cite{Rud87}). The real line $\mathbb{R}$ or suitable subsets of 
$\mathbb{R}$ are typical examples 
of the set $\Omega $ of the possible values of ${\cal A}$. 
\item The class of quantum observables represented by positive-operator-valued
measures extends the more familiar class of observables
represented by self-adjoint operators in a substantial way \cite{BGL95}.
In the special case that $\cal A$ is represented
by a self-adjoint operator $A$ on $\cal H$, 
the operator $E_{\cal A}(X)$ is simply given 
by the spectral projection of $A$ associated with $X\subseteq\Omega\subseteq
\mathbb{R} $, in 
symbols, $E_{\cal A}(X)=I_{X}(A)$. Here $I_{X}$ is the indicator
function of the set $X$. 
\item The essence of {\bf (MI)} does not depend on the formal 
frame in which quantum theory is formulated. Observables 
and states 
may be represented, as above, by operators on a Hilbert space, 
or they may, more abstractly, be postulated as elements of a 
suitable algebra and as positive linear functionals 
on this algebra, respectively. The choice of a specific mathematical 
axiomatization is irrelevant to the subsequent reasoning.
\item {\bf (MI)} does not ascribe a value of $\cal A$
to the quantum system {\em before} $\cal A$ was measured, not even in the
simple case that $\cal A$ is represented by a self-adjoint operator
with a purely discrete spectrum.
{\em After} each single measurement, however, a measurement result must 
be assigned to the individual system as a fact.
The relative frequency of the occurrence of such facts in suitable
experiments is just what the probability 
${\rm Tr}\left[ WE_{\cal A}(X)\right] $ predicts. Indeed,
the only way to interpret probabilities in physics is to compare them with
relative frequencies, that is, to interpret them statistically.
Details are given in the next section.
\end{enumerate}

\section{Reasoning}
\label{reasoning}

Our reasoning against a quantum state of the universe goes as follows.
For the physical interpretation of a quantum state in accordance with {\bf (F)}
and {\bf (MI)}, it has to be conceivable in principle to produce an
(infinite) {\em 
collection of measurement results} as facts for every observable, to ``read 
them off'' and to determine their relative frequencies. The 
interpretation of the quantum probability prediction about the observable 
$\cal A$ 
in state $\cal W$ then implies equating the probability 
${\rm Tr}\left[ WE_{\cal A}(X)\right] $
with the relative frequency of measurement results in $X$, 
for all $X\subseteq \Omega $. 
Actually, the empirical significance of $\cal W$ can be illustrated 
completely by the histograms of such collections of results for all 
observables 
that are conceivably being measured in the state $\cal W$.

Every single measurement of an observable $\cal A$ on any quantum system 
produces just one fact. 
For the empirical significance of the quantum state $\cal W$ 
it is irrelevant, whether 
different quantum systems of the same type are prepared at the same time 
into the state $\cal W$, or the same quantum 
system is repeatedly prepared into the state $\cal W$ at different times. 
In either case, an ensemble
of identically prepared quantum systems leads in the end to a collection 
of facts with the same histogram.
It is this collection of facts given {\em after} the measurements 
that serves to interpret the corresponding probability prediction
and, thereby, the quantum state. {\bf (MI)} obeys {\bf (F)}
exactly in this way. 

In order to consider the physical universe as a genuine quantum system, 
one either had to 
prepare it arbitrarily often into the same state, or one had to prepare 
arbitrarily many universes of the same type into the same state. 
In both cases, it is inconceivable in principle to register relative
frequencies of facts after measurement, supposing
the total information about the universe is encoded in its state. 

In the first case, the universe cannot remain in the same state as before a
measurement and, at the same time, exhibit the result of this measurement. 
In the second case, ``reading 
off'' relative frequencies contradicts {\bf (U)}: A universe consisting of 
{\em all} physical objects and phenomena by definition, cannot be compared 
with {\em additional} facts from ``parallel universes'', that is, from
``outside''.

Consequently, a collection of measurement results (in the sense explained
above) for the system ''universe''
cannot consistently be conceived of. 
Therefore the concept ``quantum state of the universe'' is lacking 
a sound physical interpretation,
taken for granted {\bf (F)} and {\bf (MI)}. 
Roughly speaking, any proposal to provide
this concept with empirical significance is ruled out by the probabilistic 
character of quantum theory.  
It is not enough to refute merely the more bizarre proposals 
(such as splitting the apparatus \cite{Tip86}). 
There is no way to appeal to a ``quantum state of the universe'' within
the methodological principles of physics. 

We state some obvious but far-reaching consequences of this conclusion:
\begin{enumerate}
\item There has never been a ``quantum state of the universe'' in the past. 
The origin of the physical universe cannot be explained from a 
quantum state alone, neither
by amplitudes to appear from nothing \cite{HH83} nor by a hypothetical
tunnelling phenomenon
\cite{Vil82,Lin98}.
This conclusion does not depend on whether the universe is open or closed,
inflationary or not.
There is no exclusively quantum-theoretical cosmogenesis on principle.
\item The physical universe as a whole is not subjected to a purely 
quantum-theoretical dynamics as was proposed in \cite{deW67}. 
In this sense, there is no strict quantum cosmology \cite{Haw87}.
\item A ``theory of everything'' which aims at a description of all 
physical systems and their interactions \cite{Gre99}
cannot rely exclusively upon quantum-theoretical basic 
concepts. There is no quantum theory of gravity with an interpretation 
which allows for a ``quantum state of the universe''.
\end{enumerate}

\section{Discussion}

The reasoning presented in the last section may give rise to a number 
of interesting questions, which we are going to discuss now. 
In doing so, we want to  
anticipate and refute possible objections.

{\em Question 1:} If the need for empirical accessibility is taken 
seriously, then some kind of experimental arrangement, shortly called 
``apparatus'' in the following, is indispensable. 
Does not {\em every} physical description of the universe 
necessarily comprise as {\em part} of the 
universe the apparatuses suitable to test this description? Isn't this 
problem even more fundamental than how to apply quantum theory to the 
universe as a whole? 
Isn't a {\em classical} state of the universe inconceivable as well?

{\em Answer 1:} We abstract from all concrete measurement methods. We push 
idealization even thus far as to neglect the material configuration of the 
apparatus completely. In this vein, one can relate a physical description to 
something empirically accessible in principle {\em without} explicitly
paying attention to internal states of the apparatus or to reactions 
of the apparatus to the system of interest. This stage of idealization
is well suited to find out which picture, or better caricature,
of physical reality a theory permits.
Thus, a classical {\em pure} state of the universe is conceivable 
(and has indeed 
been conceived, as is well known, in the 19th century in the guise of the 
Laplacian demon). 
Our reasoning against a quantum state of the universe notably holds true
for {\em every} probability prediction about the physical universe, be it of 
quantum origin or not. Consequently, there is also no classical {\em mixed} 
state of the 
universe, whence cosmology cannot rely on probability densities on phase space.

{\em Question 2:} In contrast to classical physics, in quantum theory state 
transformations caused by apparatuses play a central role. How can one then
justify to establish quantum descriptions 
without explicitly incorporating
preparation and measurement apparatuses?

{\em Answer 2:} All you need is {\bf (F)}. In order to interpret 
probability predictions physically, it is indispensable to 
consider collections of conceivable measurement results as facts.
Usually these facts are read off 
the apparatuses, but every description of apparatuses going beyond 
the facts themselves may fall victim to our idealization.

{\em Question 3:} Collections of measurement results can only be thought of
as produced by repeated preparation and measurement. Isn't it, in view of 
such an {\em ensemble interpretation} \cite{HW92}, 
{\em always} (and not only for the universe as defined by {\bf (U)}) 
impossible
to assign a quantum state to an individual system?

{\em Answer 3:} Whether or not a certain quantum state is 
given cannot be tested empirically in a single measurement on an 
individual system. It is, however, not {\em a priori} meaningless 
to assign a certain quantum state to an individual 
system, 
{\em as long as} one knows the preparation apparatus
(whose state is, notably, not part of the quantum description).
If it is a legitimate thought experiment to check at an 
infinite ensemble into
which quantum state a specific apparatus prepares, then it is also legitimate
to ascribe this state to each and every individual system prepared by this 
apparatus. There is no fundamental problem with this for microsystems, 
but there is one for the universe.

{\em Question 4:} The idealization relevant for interpretation extends so far
as to make irrelevant the material configuration of apparatuses ({\em
Answer 1}), as 
well as to legitimize thought experiments with infinite 
ensembles of quantum systems ({\em Answer 3}). 
Why is it then forbidden to imagine an infinite 
multitude of identically prepared quantum universes? Why should not 
different facts exist in different ``parallel universes''?

{\em Answer 4:} Abstraction and idealization 
in physics lead only to simplified
descriptions of what is empirically accessible in principle. Because  
any view from outside the universe is inconceivable by the
definition {\bf (U)}, 
a multitude of universes or a comparison between different 
universes remains forbidden, even if idealization is pushed to the extreme. 
It is legitimate to imagine an infinite ensemble of electrons, only {\em 
because} it is conceivable in principle to prepare many electrons (or one
electron repeatedly) into the same state. For the universe, the situation 
is fundamentally different.
Even if the {\em material} configuration of apparatuses 
is completely neglected,
there remains a difference in their {\em logical} status: An apparatus for the 
observation of an electron is surely outside the electron, but an apparatus
for the observation of the universe is surely {\em not} outside the universe.

{\em Question 5:} 
The real structure of nature does not depend on definitions. {\em Answer 4}, 
however, seems to do so. Why are cosmological scenarios excluded which
involve a multitude of universes, {\em each} being part of nature?
``Universe'' means
``all embracing'', but why should a physical universe as an object
of cosmology be literally everything?

{\em Answer 5:} One can, of course, give up {\bf (U)} and use the word 
``universe'' in a less embracing sense. But then, our reasoning 
and conclusion remain valid for what was originally meant by {\bf (U)}.

{\em Question 6:} Real apparatuses consist of atoms, and atoms are
undisputedly quantum systems. Why can one then rely on facts to interpret 
quantum states without describing the emergence of these facts 
within the conceptual frame of quantum theory? 
Doesn't the whole reasoning rest on an artificial opposition between
{\em quantum} predictions and {\em classical} apparatuses due to 
over-idealization, and hence lack physical relevance?

{\em Answer 6:} Indeed, the application of {\bf (MI)} {\em presupposes} 
that a collection of facts comes out of every sequence of measurements.
{\bf (MI)} gives no hints on how these facts come 
into existence or on how their emergence could be described theoretically.
This notorious ``quantum measurement problem'' \cite{WZ83} 
cannot be solved or avoided by explicitly taking into account 
the apparatus {\em and} the environment 
as quantum systems. In particular, purely quantum-dynamical 
theories of decoherence \cite{Giu96} do not explain the emergence of facts
in single measurements,
not even for all practical purposes.
The idealization chosen here favours 
the sudden emergence of a definite fact in a spontaneous quantum event 
\cite{Haa96} once a measurement is carried out on a quantum system.
From then on the fact persists. Conventional quantum theory expresses 
such an individual quantum event as a suitable state collapse.
Encouraged by these facts and in the tradition of Niels Bohr, 
we insist that the classical 
description of apparatuses is a necessary independent input to every
quantum description. By ``classical'' we do not refer to the laws
of classical physics, but only to the applicability of classical logic to the
facts presupposed by {\bf (MI)}. The relevance of these facts to 
interpretation is a direct consequence of {\bf (F)}.

{\em Question 7:} The unsatisfactory special role of the apparatuses and the
desire for a description of the universe as a closed quantum system have been
two essential motivations for the development of the formalism of {\em 
consistent quantum histories} \cite{Gri84}. 
Hasn't the state concept lost its fundamental status in this 
modification of the quantum formalism, so that the reasoning presented above
has become obsolete?

{\em Answer 7:} 
Quantum probability goes without histories, mathematically \cite{Mey95}
and physically.
The formalism of consistent quantum histories is burdened with a fundamental
freedom of choice of a consistent family or a framework \cite{Gri96}.
Among the various imaginable quantum histories, there is no unique 
procedure to discriminate in a given physical situation between facts
and non-facts. In the histories formalism it is not unambiguously
expressible that one observable has an actual value due to the real 
experimental setup, while another (incompatible) one has not. 
This is the ultimate reason why the histories
approach has been criticized repeatedly \cite{Esp87}. Independently of 
its applicability to the universe, the quantum-histories approach thus
fails to satisfy principle {\bf (F)}. For this reason it lacks a 
sound physical interpretation. 
This is fatal to the whole approach, but it is far from
being acknowledged by its adherents \cite{Gri98}.

\vspace{0.5cm}

Finally, one could ask how to do cosmology at all in the era of quantum 
theory. We stress that this problem appears to be puzzling only 
through the dogma of the {\em universal} applicability of quantum theory. 
In accordance with Ludwig \cite{Lud85} and others, we suggest to drop this
dogma. In the same way
as the description of a quantum-mechanical microsystem requires classical
apparatuses as a fundamental concept, facts could come into play in the 
description of the early universe and within grand-unification programs,
as a fundamental concept apart from quantum uncertainty. We think that this 
dichotomy is unavoidable. Moreover, it is by no means 
evident that all physical systems must possess quantum states.

In conclusion, we have shown that the universe as a whole 
cannot be ascribed a quantum state with a sound interpretation,
irrespectively of specific cosmological models. 
Thus, it makes 
no sense to postulate such a state hypothetically and treat
it like a very complicated quantity, about which one doesn't yet know enough.
This conclusion should serve as an interpretational boundary condition 
for working out cosmological theories. Its enforcing character is 
based on conceptual and methodological rigor. 
This is a step beyond Occam's razor, which has so often been the main tool
of heuristic argumentation against a multitude of universes: While
the razor cuts off only what is physically legitimate but redundant, 
the idea of an ensemble of universes is at best metaphysical.


\end{document}